\documentclass[twocolumn,showpacs,preprintnumbers,amsmath,amssymb,prb,aps]{revtex4-1}
\usepackage{subfigure}
\usepackage{esvect}
\usepackage[usenames, dvipsnames]{color}
\usepackage{comment}
\usepackage{epsfig}% Include figure files
\usepackage{graphicx}% Include figure files
 \usepackage{float} % for pinning a figure
\usepackage{gensymb}% for degree
\usepackage[colorlinks=true,linkcolor=red,citecolor=blue,urlcolor  = blue]{hyperref}
\usepackage{graphicx}  % needed for figures
\usepackage{dcolumn}   % needed for some tables
\usepackage{bm}        % for math
\usepackage{ulem}
\begin{document}

%Title of paper
\title{Structural anomaly in superconductivity of CaFe$_2$As$_2$ class of materials}

\author{Ram Prakash Pandeya$^1$, Arindam Pramanik$^1$, Anup Pradhan Sakhya$^1$, Rajib Mondal$^1$, A. K. Yadav$^2$, S. N. Jha$^2$, A. Thamizhavel$^1$ and Kalobaran Maiti$^1$}
\altaffiliation{Corresponding author: kbmaiti@tifr.res.in}

\affiliation{$^1$ Department of Condensed Matter Physics and Materials Science, Tata Institute of Fundamental Research, Homi Bhabha Road, Colaba, Mumbai - 400 005, INDIA. \\
$^2$ Nuclear Physics Division, Applied Spectroscopy Division, Bhabha Atomic Research Centre, Mumbai - 400 085, INDIA.}

\date{\today}

\begin{abstract}
Quantum transitions in Fe-based systems are believed to involve spin, charge and nematic fluctuations. Complex structural phase diagram in these materials often emphasizes importance of covalency in their exotic properties, which is directly linked to the local structural network and barely understood. In order to address this outstanding issue, we investigate the evolution of the structural parameters and their implication in unconventional superconductivity of 122 class of materials employing extended $x$-ray absorption fine structure studies. The absorption spectral functions near the Fe $K$-edge and As $K$-edge of CaFe$_2$As$_2$ and its superconducting composition, CaFe$_{1.9}$Co$_{0.1}$As$_2$ ($T_c$ = 12 K) exhibit evidence of enhancement of Fe contributions near the Fermi level. As-Fe and Fe-Fe bondlengths derived from the experimental data show interesting changes with temperature across the magneto-structural transition; the evolution of these parameters in Co-doped composition is similar to its parent compound despite absence of magneto-structural transition in the studies of their bulk properties. These results reveal evidence of doping induced evolution to the proximity to critical behavior presumably leading to superconductivity in the system.
\end{abstract}

\maketitle

\section{Introduction}

While high temperature superconductivity continues to be an outstanding puzzle in contemporary condensed matter physics, the Fe-based systems revealed additional complexity in the problem. It appears that the complex interplay of spin, orbital, charge and lattice degrees of freedom is responsible for the exoticity of these materials.\cite{HHosono2015} Parent compound of almost all the Fe-based materials exhibit a structural transition from tetragonal phase possessing C4 symmetry to an orthorhombic phase, where the 4-fold rotational symmetry is broken leading to nematicity in the system.\cite{HHosono2015,Stewart} Some studies pointed out importance of charge reservoir layer in the electronic properties.\cite{THGeballe,Khadiza-DFT} Electronic properties of these systems have been studied extensively using angle-resolved photoemission spectroscopy (ARPES), scanning tunnelling spectroscopy, etc.\cite{STan,CLiu,ACharnukha,Ganesh-Eu122,Hoffman-STM} However, the structural aspects remains to be still illusive, in particular, the role of local structural parameters in the electronic properties.

In order to address this issue, we employed extended $x$-ray absorption fine structure (EXAFS) studies to probe the evolution of structural parameters of a archetypical compound CaFe$_2$As$_2$ (Ca122), a parent Fe-based superconductor in the 122 class of materials and its Co-doped composition, CaFe$_{1.9}$Co$_{0.1}$As$_2$ (CaCo122), which shows superconductivity below 12 K. Ca122 has been studied extensively and shows both structural and magnetic transition from tetragonal paramagnetic phase to orthorhombic-antiferromagnetic phase below 170 K.\cite{NNi} In addition, it shows complex phase diagram as a function of pressure; application of a small pressure ($\sim$0.35 GPa) helps to retain its tetragonal symmetry, which is called a collapsed tetragonal (cT) phase.\cite{Kreyssig,Goldman,Pratt,Ca122-Ram,Milton} The cT phase can be obtained by other means too such as chemical substitution at any of the three atomic cites and/or quenching from a temperature higher than 700 $^o$C to room temperature.\cite{ARPES-Dhaka} Even the ambient pressure phase shows signature of cT phase in its electronic structure.\cite{Khadiza-ARPES} Many studies showed superconductivity on application of pressure and/or doping, \cite{Milton,YQi,Saha,KKudo,Chen,KZhao} while some other studies did not find superconductivity in the cT phase.\cite{WYu} Evidently, superconductivity in this system is complex along with additional complexity arising due to structural and magnetic interactions. Here, we report intriguing results in Ca122 and CaCo122 systems revealing an interesting structural link to superconductivity of this material.

\section{Experimental Technique}

Single crystalline samples of CaFe$_2$As$_2$ and CaFe$_{1.9}$Co$_{0.1}$As$_2$ were grown using Sn flux \cite{NKumar,RMittal} and characterized by powder \textit{x}-ray diffraction and energy dispersive analysis of \textit{x}-rays measurements. We carried out magnetic susceptibility measurements on both these samples and observe signature of antiferromagnetic transition at 170 K in Ca122.\cite{NNi} However, no such transitions are observed in the magnetic measurements of the doped composition, CaCo122. A sharp diamagnetic transition below 12 K is observed in CaCo122 indicating its superconducting phase in the ground state.\cite{RMittal}

EXAFS measurements were carried out at BL-8 beamline at INDUS-2, RRCAT, Indore, India in transmission mode using the samples in thin pellet form; sample was ground together with cellulose powder to make homogeneous pellets.\cite{BL8_Indus2} The measured data were processed using \textit{Athena} software and the structural parameters were extracted using \textit{FEFF} and \textit{FEFFIT} packages as implemented in \textit{Artemis} software.\cite{Software}

\section{Results and discussion}

\begin{figure}
   \centering
   \includegraphics[width = 1.0\linewidth]{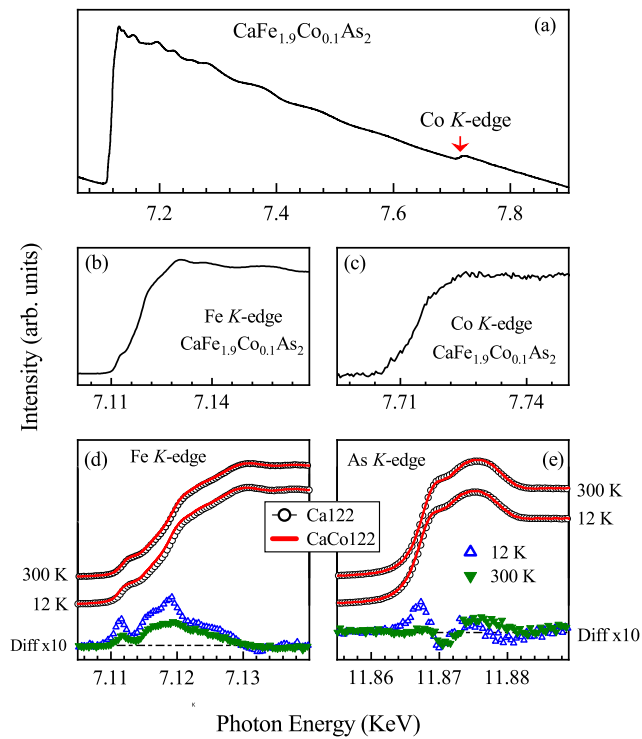}
\vspace{0ex}
\caption{(a) Fe $K$-edge absorption spectrum of CaCo122 collected at 300 K. Co $K$-edge is shown by an arrow. Expanded view of (b) Fe $K$ -edge and (c) Co $K$-edge regimes. Comparison of the (d) Fe $K$-edge and (e) As $K$-edge data taken from Ca122 (open circles) and CaCo122 (line) at 300 K and 12 K. Difference spectra (CaCo122 data - Ca122 data) rescaled by 10 times are shown at the bottom panel for the data at 12 K (open triangles) and 300 K (solid triangles).}
 \label{Fig1:XANES}
\end{figure}

In Fig. \ref{Fig1:XANES}(a), we show Fe $K$-edge and Co $K$-edge $x$-ray absorption spectra of CaCo122 collected at 300 K. The near edge regimes, shown in expanded energy scales in Fig. \ref{Fig1:XANES}(b) and Fig. \ref{Fig1:XANES}(c), exhibit similar features indicating similarities in the transition metal 4$p$/3$d$ contributions in the unoccupied part of the electronic structure. The As $K$-edge data represent transition to the unoccupied As 4$p$ partial density of states (PDOS) and possess Fe 3$d$ symmetry due to strong Fe 3$d$-As 4$p$ covalency.\cite{Khadiza-DFT} The Fe $K$-edge results represent the unoccupied part of the Fe 4$p$ PDOS, which also contains Fe 3$d$ symmetry due to finite Fe 4$p$-3$d$ hybridization. In order to probe the electronic structure near the absorption edge, the Fe $K$-edge and As $K$-edge spectra of Ca122 and CaCo122 are compared in Fig. \ref{Fig1:XANES}(d) and Fig. \ref{Fig1:XANES}(e) after normalization by the intensities at about 40 eV away from the absorption edge. The energies of the experimental spectral features appear very close to those found in the EXAFS database for Fe and As elemental metals consistent with the studies in other Fe-based systems \cite{Ba122-EXAFS} as well as in the hard \textit{x}-ray photoelectron spectroscopic studies of Ca122.\cite{Ca122-Ram}

We have superimposed the Ca122 and CaCo122 data in Fig. \ref{Fig1:XANES}(d) and Fig. \ref{Fig1:XANES}(e) to investigate the changes due to Co substitution at 300 K and 12 K. While the raw data look very similar, the difference spectra manifest interesting features. For example, As $K$-edge data at 300 K shown in Fig. \ref{Fig1:XANES}(e) exhibits a weak spectral weight transfer to higher energies due to Co substitution suggesting change in As 4$p$-Fe 3$d$ hybridization at 300 K to be small. However, an intense peak appears near the absorption edge of the difference spectrum at 12 K, while the higher energy regimes remain almost the same. The difference spectrum at Fe $K$-edge reveals an enhancement in intensity with two distinct peaks; a narrower one at the edge and a broader one at higher energy. Co-substitution at the Fe-sites dopes electrons and therefore, the absorption edge is expected to shift to higher energy and/or a change occurs in the charge states of the atoms within the quasi-2D FeAs layers. However, the features near the edge in Fig. \ref{Fig1:XANES}(d) indicate a small shift towards lower energies. One possible reason for such an anomaly could be a somewhat stronger Fe 4$p$-3$d$ hybridization in the doped compound leading to an enhancement of relative Fe 4$p$ population near the Fermi level. At 12 K, the intensities of the features near the edge enhances significantly relative to the higher energy ones.

\begin{figure}
   \centering
   \includegraphics[width = 0.95\linewidth]{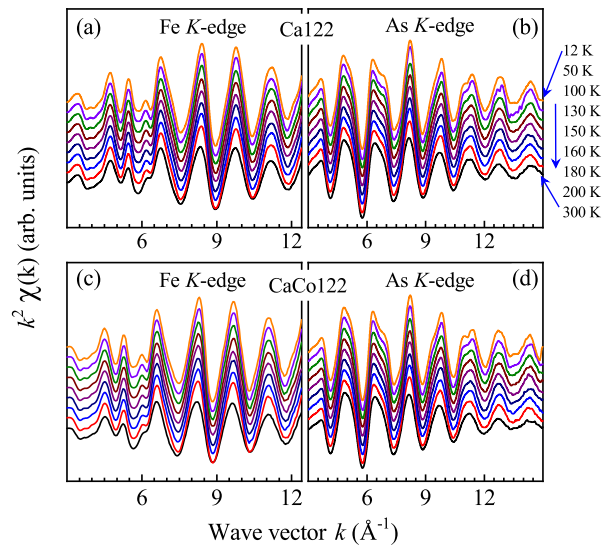}
\vspace{-1.5ex}
\caption{$k^2$ weighted EXAFS oscillations extracted from (a) Fe $K$-edge and (b) As $K$-edge data of Ca122. Similar results for CaCo122 at (c) Fe $K$-edge and (d) As $K$-edge.}
\label{Fig2:ChiK}
\end{figure}

The extracted $k^2$-weighted EXAFS oscillations in the Fe and As \textit{K}-edge spectra collected on Ca122 and CaCo122 samples are shown in the upper and lower panels of Fig. \ref{Fig2:ChiK}, respectively. It appears that Co-doping in Ca122 does not have distinctly observable change in the data except a weak change in intensity of the peaks at higher $k$-values. The data as a function of temperature, however, exhibit interesting evolution; signatures of distinct evolution are clearly visible at higher $k$-values. For example, in the As $K$-edge EXAFS oscillations, we observe a gradual merging of the two nearby features at the $k$-value of about 11\AA$^{-1}$ when the temperature is increased. The Fe $K$-edge data show broadening of the peaks with the enhancement of temperature as expected. Clearly, there is significant change in structural parameters as a function of temperature.

\begin{figure}
   \centering
   \includegraphics[width = 0.95\linewidth]{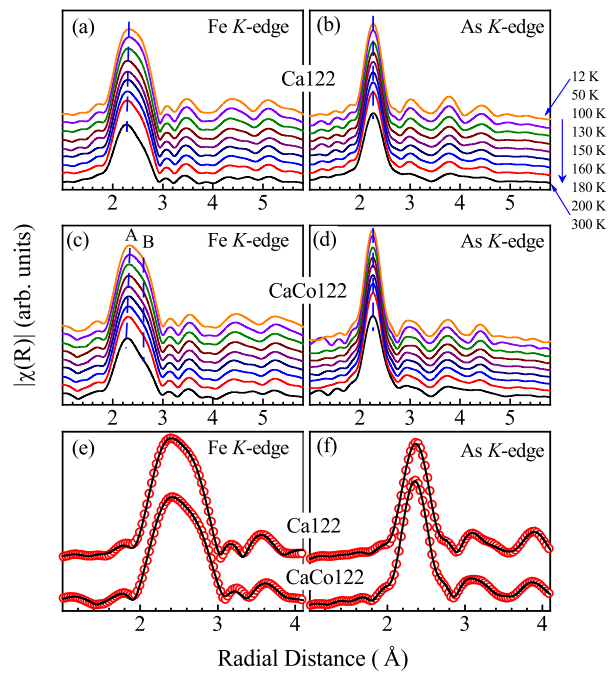}
\vspace{0ex}
\caption{Fourier transform of the temperature dependent $k^2$ weighted EXAFS oscillations extracted from (a) Fe $K$-edge and (b) As $K$-edge data of Ca122. Similar results for CaCo122 at (c) Fe $K$-edge and (d) As $K$-edge. Theoretical simulation of FTs of (e) Fe K-edge and (f) As K-edge EXAFS oscillations measured at 12 K.}
\label{Fig3:ChiR}
\end{figure}

The effects observed in EXAFS oscillations can be more precisely captured in the Fourier transform (FT) of oscillations in real space as shown in Fig. \ref{Fig3:ChiR}. The FT of the $k^2$-weighted EXAFS oscillations were calculated using a Hanning window of 3 - 15 \AA$^{-1}$ in the $k$-axis; due to the presence of Co $K$-edge in Fe $K$-edge spectra of CaCo122, the window was kept within 3-12.5 \AA$^{-1}$ for Fe $K$-edge data analysis. FTs corresponding to the Fe $K$-edge and As $K$-edge data of Ca122 are shown in Fig. \ref{Fig3:ChiR}(a) and Fig. \ref{Fig3:ChiR}(b), and the same for CaCo122 sample are shown in Fig. \ref{Fig3:ChiR}(c) and Fig. \ref{Fig3:ChiR}(d), respectively; black to orange lines correspond to the spectrum collected at the temperatures 300 K to 12 K. The intensity and the peak position of the features in the As $K$-edge data of both the samples show interesting evolution with temperature. The most intense peak in the Fe $K$-edge data exhibit two peak feature (A and B in the figure) and a shift towards lower values with the increase in temperature indicating compression instead of expansion expected. The relative intensity of the feature, B reduces significantly with the increase in temperature.

To extract the parameters related to the local structure, the radial distribution of FT amplitudes shown in Fig. \ref{Fig3:ChiR}, were modeled by the conventional approach based on single scattering approximation. The crystal structure and atomic positions in Ca122 are well studied employing high resolution $x$-ray and neutron diffraction,\cite{NNi,Kreyssig,Goldman} which is used as a guide for the analysis of the EXAFS oscillations. Since the distortion due to structural transition is small, we report the results corresponding to effective tetragonal structure for clarity in presentation.

As atoms are surrounded by a tetrahedral cage formed by four Fe atoms (FeAs bond length $\sim$2.4 \AA) and the first shell of As $K$-edge data contains contribution from these As-Fe bonds. However, the first peak in the FT of Fe $K$-edge data contains both Fe-As and Fe-Fe contributions as evident in the Fig. \ref{Fig3:ChiR}(a) and \ref{Fig3:ChiR}(c); the difference between Fe-Fe and Fe-As distances are not enough to get distinct features. Therefore, the local structure information related to the Fe-As bonds is extracted from As $K$-edge, which is used to extract the Fe-Fe bond parameters from the Fe $K$-edge data.\cite{BJoseph,Hacisalihoglu} In order to find the Ca-$X$ ($X$ = Fe, As) bondlengths, the simulation was carried out up to the second order shell as done earlier for similar systems.\cite{Paris} The simulated results obtained for Fe and As $K$-edge data at 12 K for both the samples are shown in Fig. \ref{Fig3:ChiR}(e) and \ref{Fig3:ChiR}(f) manifesting good representation of the experimental results.

\begin{figure}
   \centering
   \includegraphics[width =0.95\linewidth]{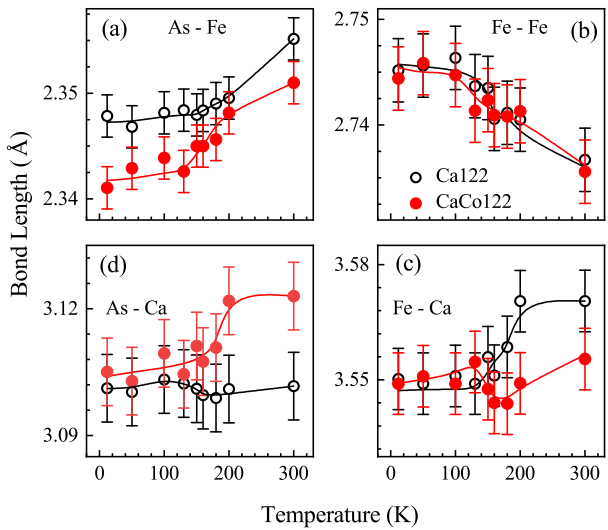}
\vspace{0ex}
\caption{(a) As-Fe, (b) Fe-Fe, (c) As-Ca and (d) Fe-Ca bondlengths as a function of temperature in Ca122  (open circles) and CaCo122 (solid circles). Solid lines are the smooth curves drawn as a guide to the eye. Bond distances with Ca exhibit interesting temperature evolution.}
\label{Fig4:bond}
\end{figure}

Extracted bondlengths are shown in Fig. \ref{Fig4:bond} exhibiting intriguing temperature dependence. The results for the parent compound, Ca122 shown by open circles show significant change in bond distances across the magnetic and structural phase transition temperature of 170 K. The As-Fe bondlength gradually decreases from the room temperature across the magneto-structural transition and eventually stabilizes at lower temperatures. Instead, the Fe-Fe distance increases gradually and saturates below the transition temperature. These results reveal decrease in anion height with the decrease in temperature and hence, stronger Fe-As hybridization along with a decrease in Fe-Fe direct overlap within the $xy$-plane. Ca layers show an unusual scenario; while the As-Ca distance marginally increases across the phase transition, Fe-Ca distance reduces significantly with the decrease in temperature indicating stronger overlap of Ca 4$s$ states with the Fe 3$d$ states. This suggests an important role of Ca-sites as found in other systems, they are not just spectator elements in the structure.\cite{Debdutta-ruthenates} The thermalization of the bondlengths appears to occur at a temperature of about 100 K, which is much below the transition temperature.

In CaCo122, the As-Fe distance becomes smaller than that in the parent compound. Although no magnetic transition is observed in this material, we observe a significant dip spreading over a wider temperature range at the low temperature side. Fe-Fe distance exhibits behavior similar to Ca122. As-Ca distance is larger in the doped sample than the distance in the pristine one. Below 200 K, we observe a stiff decrease in the As-Ca distance, which becomes close to the values in the pristine sample below 100 K. Fe-Ca distance manifest anomalous evolution - at room temperature, the bondlength is smaller in the doped sample and gradually reduces with the temperature as expected. Below 200 K, it exhibits small sudden jump to a value close to room temperature value and saturates below 100 K. Clearly, onset of magnetic interactions at low temperature induces significant structural reorganization and a new structural order seems sets in below 100 K. The structural transition in Ca122 leads to a reduction of rotational symmetry from C4 (tetragonal) to C2 (orthorhombic) phase which is defined as nematicity. Recent studies revealed signature of hidden C4 phase (collapsed tetragonal phase) even in the parent compound at ambient pressure probably due to strain.\cite{Khadiza-ARPES} Evidently, Co-substitution in Ca122 brings the system closer to its magneto-structural instabilities (an enhanced nematic fluctuations) and absence of magnetic order although magnetic interactions are present as in Ca122, which is a signature of proximity to quantum criticality. Interestingly, Ca appears to play an important role in deriving the electronic properties.

\begin{figure}
   \centering
   \includegraphics[width = 0.95\linewidth]{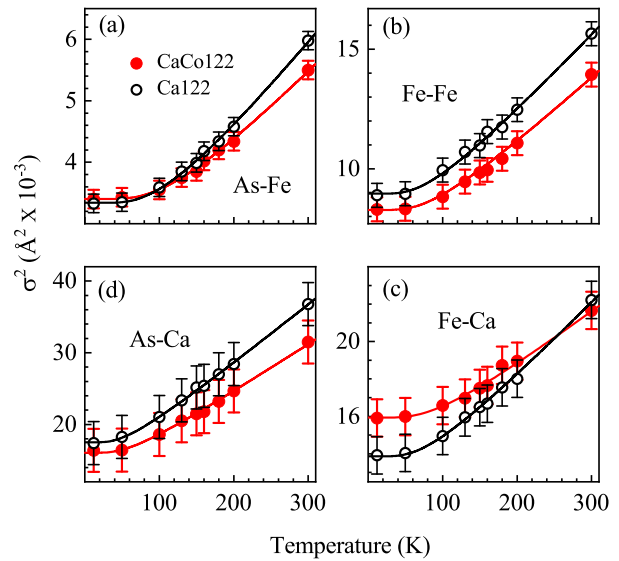}
 \vspace{0ex}
\caption{Debye-Waller factor corresponding to (a) As-Fe, (b) Fe-Fe, (c) Fe-Ca and (d) As-Ca bonds in Ca122 (open circles) and CaCo122 (solid circles). Lines are the fit employing Einstein model.}
\label{Fig5:DWF}
\end{figure}

We estimated the Debye-Waller factor (DWF), $\sigma^2$ associated to each pair of atoms; the results (symbols) are shown in Fig. \ref{Fig5:DWF}. In every case, DWF gradually reduces with the decrease in temperature and gets saturated at low temperatures. This typical behavior is in line with the expected behavior arising from thermal contributions. We do not observe any anomaly across the structural and magnetic phase transition temperature unlike other correlated systems such as manganites.\cite{Bindu-PRB} The DWF for As-Fe is found to be the smallest (strong bonding) and that for As-Ca is the highest (weak bonding). We observe an enhancement of DWF for Fe-Ca with Co substitution as expected,\cite{BJoseph2} while it is very similar for As-Fe bonds. Curiously, DWF for Fe-Fe and As-Ca bonds becomes smaller with Co-doping although the temperature dependence look very similar in both the cases.

Using Einstein model of crystal vibration, DWF is given by, $\sigma^2 = \sigma_\circ^2 + (\hbar/2 \mu \omega_E)  coth[(\hbar \omega_E)/(2 k_B T)]$, where $\mu$, $\omega_ E$ and $\sigma_\circ^2$ are the reduced mass, the Einstein frequency and temperature independent part of DWF, respectively.\cite{DW-factor} For a given pair of atoms of mass $m_1$ and $m_2$, the reduced mass, $\mu$  (= $m_1m_2/(m_1+m_2)$) is independent of temperature and $\sigma_\circ^2$ is linked to the lattice disorder. The temperature dependent part is related to the thermodynamic properties of the material and allows us to derive the Einstein temperature, $\Theta_E$ (= $\hbar\omega_E/k_B$) and the effective spring constant, $k_{eff}$ (= $\mu\omega_E^2/2$). The lines passing through the symbols in the Fig. \ref{Fig5:DWF} represent the least square error fits and provide an excellent representation of the experimental results within the Einstein model.

\begin{table}
\caption{Local structure parameters obtained from temperature dependent of Fe and As $K$-edge EXAFS measurements on CaFe$_2$As$_2$ (CaFe$_{1.9}$Co$_{0.1}$As$_2$).}\label{Ca122}
\begin{ruledtabular}
\begin{tabular}{lcccc}
 Bonds& $\Theta_E$ (K) & $\omega_{E}$ (meV)& $k$ (eV.\AA$^{-2}$) & $\sigma_\circ^2$  (\AA$^2$) \\
\hline
  Fe-As & 310 (337) & 25.8 (28.0) & 5.1 (6.0) & 0.001 (0.001) \\
  Fe-Fe & 231 (249) & 19.2 (20.4) & 2.5 (2.8) & 0.005 (0.005) \\
  Fe-Ca & 228  (265) & 19.0 (22.0) & 2.0 (2.7) & 0.009 (0.012) \\
  As-Ca & 150(167) & 12.5 (13.9) & 1.0 (1.2) & 0.011 (0.011) \\
\end{tabular}
\end{ruledtabular}
\label{tab:Ca122}
\end{table}

The extracted parameters for Ca122 and CaCo122 are given in the Table. \ref{tab:Ca122}. In Ca122, the Einstein temperature, $\Theta_E$ is 310 K (215 cm$^{-1}$) for Fe-As bond, which is very close to the measured Raman shift in Ca122 for $E_g$ mode (211 cm$^{-1}$); Fe and As atoms oscillate opposite to each other in the respective $ab$ planes.\cite{Raman-Choi} The Einstein frequency and spring constant is the largest for Fe-As bond; these results provide an independent evidence of strong Fe 3$d$ - As 4$p$ covalency. Similar spring constants for Fe-Ca bonds in comparison to Fe-Fe bonds, suggests importance of Fe 3\textit{d}-Ca 4\textit{s} hybridization. It is the smallest for As-Ca bond (150 K) indicating weaker but non-negligible bonding along with a larger static disorder, $\sigma_\circ^2$.

With Co-substitution, the Einstein temperature, Einstein frequency and spring constant enhances significantly keeping the static disorder close to the values in parent compound as observed in other Fe-based superconductors.\cite{BJoseph2,Hacisalihoglu,Granado} Significant increase in spring constants for Fe-As and Fe-Ca bonds with Co doping suggests strengthening of Fe 3$d$-As 4$p$ and Fe 3$d$-Ca 4$s$ hybridizations in the doped compound. It appears that the parameters (spring constant, Einstein frequency, {\it etc.}) associated to the Fe-sites play a major role in the properties of the superconducting composition.  This has also been manifested in the near-edge data, where we observed an enhancement of overall Fe contributions at the Fermi edge and a shift of the absorption edge towards lower energies due to Co-substitution.

It is to be noted here that a larger $k$-range is desirable for the estimation of the parameters associated to Ca-X bonds, which is not possible in the present case due to the presence of Co $K$-edge. Nevertheless, we make a comment on the estimated parameters based on the available data. In the Table. \ref{tab:Ca122}, we observe that the values of $\sigma_\circ^2$ obtained for both As-Ca and Fe-Ca bonds are quite high indicating relatively large disorder at Ca sites. Recent angle-resolved photoemission study \cite{Khadiza-ARPES} revealed signature of collapsed tetragonal phase hidden even within the ambient band structure, which correspond to a slightly smaller lattice constant, $c$ and subsequent enhancement of lattice parameters in the $ab$-plane. Thus, one reason for larger $\sigma_\circ^2$ could be the presence of this hidden structure. Overall, the presence of larger disorder associated to Ca-site maintaining other cases small indicate that the disorder arising due to the chemical substitutions are presumably absorbed by the Ca-layer as expected for a charge-reservoir layer in such materials.

\section{Conclusion}

In summary, we have studied the role of structural parameters in the electronic properties of Ca122 with an emphasis on their evolution in achieving superconductivity. Fe contribution near the Fermi level appear to be somewhat larger in the superconducting composition relative to the parent compound. In CaFe$_2$As$_2$, the bondlengths gradually changes over a wide temperature range; the change in parameters starts at a temperature much higher than the magneto-structural transition temperature of 170 K providing evidence of a precursor effect.\cite{RSSingh,Sampath} A flattening of the bondlengths occurs below 100 K indicating achieving the structural parameters close to its ground state configuration. Curiously, similar behavior is observed in the Co-doped sample too which is superconducting and does not show magnetic transition. Disorder appears to be similar in both the cases despite the chemical substitution in the superconducting material. The evidence of the survival of electronic interactions of the parent compound within the superconducting composition suggest proximity of the superconducting material to quantum criticality as found in other systems.\cite{ruthenates}

\section{Acknowledgements}

Authors acknowledge RRCAT, Indore for their support in carrying out experiments and financial support under the project no. 12-R\&D-TFR-5.10-0100. KM acknowledges financial assistance from DST, Govt. of India under J. C. Bose Fellowship program and DAE under the DAE-SRC-OI Award program.

\end{document}